\documentclass[]{spie}  
\pdfoutput=1
\usepackage{amsmath,amssymb}
\usepackage[]{graphicx}

\def\bicep{{\sc Bicep}}

\def\acbar{{\sc Acbar}}
\def\QUAD{{\sc QUaD}}

\def\boomn{{\sc Boomerang}}

\def\muK{~\mu{\rm K}}

\def\deg{^\circ}

\title{CMB polarimetry with BICEP: \\
instrument characterization, calibration, and performance}

\author{Yuki D. Takahashi*\supit{a}, Denis Barkats\supit{b},
John O. Battle\supit{c}, Evan M. Bierman\supit{d}, James J. Bock\supit{b,c}, \\
H. Cynthia Chiang\supit{b}, C. Darren Dowell\supit{c}, 
Eric F. Hivon\supit{e}, William L. Holzapfel\supit{a}, \\
Viktor V. Hristov\supit{b}, William C. Jones\supit{b}, J. P. Kaufman\supit{d},
Brian G. Keating\supit{d}, John M. Kovac\supit{b}, \\
Chao-Lin Kuo\supit{f}, Andrew E. Lange\supit{b}, Erik M. Leitch\supit{c},
Peter V. Mason\supit{b}, Tomotake Matsumura\supit{b}, \\
Hien T. Nguyen\supit{c}, Nicolas Ponthieu\supit{g}, 
Graca M. Rocha\supit{b}, Ki Won Yoon\supit{h},
P. Ade\supit{i}, L. Duband\supit{j}
\skiplinehalf
\supit{a} University of California, Berkeley, USA; \\
\supit{b} California Institute of Technology, Pasadena, USA; \\
\supit{c} Jet Propulsion Laboratory, Pasadena, USA; \\
\supit{d} University of California, San Diego, USA; \\
\supit{e} Institut d'Astrophysique de Paris, France; \\
\supit{f} Stanford University, Palo Alto, USA; \\
\supit{g} Universite Paris XI, Orsay, France; \\
\supit{h} National Institute of Standards and Technology, Boulder, USA; \\
\supit{i} Cardiff University, UK; \\
\supit{j} Commissariat \`{a} l'\'{E}nergie Atomique, Grenoble, France; \\
}

\authorinfo{* yuki @ bolo.berkeley.edu; phone 1 510 642 4359; fax 1 510 643 5204}

\begin{document}
  \maketitle

\begin{abstract}
\bicep\ is a ground-based millimeter-wave bolometric array designed to 
target the primordial gravity wave signature on the polarization of 
the cosmic microwave background (CMB) at degree angular scales.  Currently 
in its third year of operation at the South Pole, \bicep\ is measuring the 
CMB polarization with unprecedented sensitivity at 100 and 150 GHz in the 
cleanest available 2\% of the sky, as well as deriving independent 
constraints on the diffuse polarized foregrounds with select observations 
on and off the Galactic plane.  Instrument calibrations are discussed in 
the context of rigorous control of systematic errors, and the performance 
during the first two years of the experiment is reviewed.
\end{abstract}

\keywords{cosmic microwave background polarization, mm-wave, bolometers, 
cosmology, inflation, South Pole}

\section{INTRODUCTION}

The clearest evidence of an inflationary origin of the Big Bang would be 
a detection of a curl component (``B-mode") polarization of the CMB arising 
from gravity wave perturbations at the time of CMB decoupling.
The B-mode polarization is expected to peak at degree angular scales, with 
the magnitude of the power spectrum described by the ratio $r$ of the 
initial tensor to scalar perturbation amplitudes, a quantity directly 
related to the energy scale of inflation.  
The current upper limit is $r$$<$0.2 from CMB temperature 
measurement at large angular scales by WMAP, combined with 
constraints from Type Ia supernovae and baryon acoustic oscillations 
\cite{2008arXiv0803.0547K}.
More directly, upper limits on B-mode polarization of $\sim$0.8 
$\mu$K rms have been placed by WMAP at multipole of $\ell$$\sim$65 
and by \QUAD\ at $\ell$$\sim$200, respectively \cite{2008arXiv0805.1990Q}.
These limits are still well above the expected levels of confusion from 
the Galactic synchrotron and dust foregrounds in the cleanest
regions of the sky and from lensing which converts the much brighter 
gradient component (``E-modes") to B-modes at smaller angular scales.
An instrument designed to target the expected peak of the gravity-wave 
signature at $\ell$$\sim$100, with judicious selection of observed field 
and exquisite control of systematic errors, should be sensitive to 
values of $r$ well below the current upper limit.  Using proven bolometric 
technologies albeit with unprecedented total sensitivity, \bicep\ is just 
such an effort.  
Reaching a sensitivity corresponding to $r$=0.1, or a signal at degree scales of 
$\sim$0.1 $\mu$K rms, requires careful instrument characterization and 
calibration to minimize systematic contamination in the polarization 
measurement.

\section{OVERVIEW OF INSTRUMENT AND OBSERVING STRATEGY}

\begin{figure}[!tb]
 \begin{center} \begin{tabular}{c}
 \includegraphics[width=\linewidth]{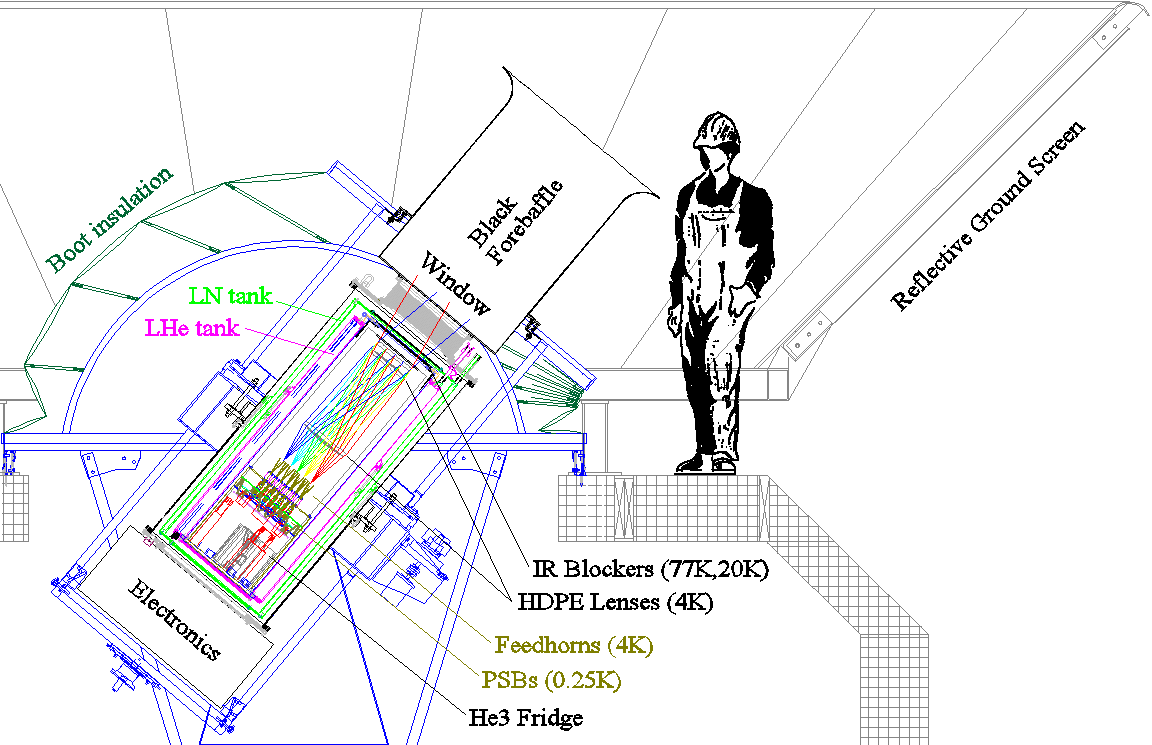}
 \end{tabular} \end{center}
 \caption[instrument]{\label{fig:instrument} \bicep\ telescope at the 
South Pole, at its lowest observing elevation of 50$\deg$.
The LN$_2$/LHe toroidal cryostat encloses the entire 4K telescope, and 
houses a $^4$He/$^3$He/$^3$He sorption refrigerator that 
cools the bolometers to 250 mK.}
\end{figure}

Installed at the South Pole in November 2005, \bicep\ is a compact 
refractor with 49 pairs of polarization sensitive bolometers (PSBs) at 
primarily 100 GHz and 150 GHz with $0.9\deg$ and $0.6\deg$ beams, 
respectively.  
The bolometers use neutron transmutation doped (NTD) Germanium 
thermistors to sense the optical power incident on the absorber mesh.  
Figure~\ref{fig:instrument} shows a schematic drawing of the instrument,
looking out through the roof of the Dark Sector Laboratory (DSL) building 800 
meters from the geographic pole at 2800-meter altitude. 
The instrument design and the observation strategy were described in Yoon 
et al. \cite{Yoon2006} (hereafter referred to as the ``2006 Paper").

\begin{figure}[tbp]
\begin{minipage}[t]{0.45\linewidth}
\small \begin{center} 
\begin{tabular}[c]{|l|cc|}
\hline & 2006 & 2007/2008 \\
\hline
100 GHz & 25 (19) & 25 (22) \\
150 GHz & 24 (14) & 22 (15) \\
220 GHz & 0 & 2 \\
dark bolometers & 6 & 6 \\
thermistors (NTD) & 6 & 8 \\
resistors (5 M$\Omega$) & 6 & 5 \\
\hline
\end{tabular}
\end{center}
   \caption[summary]
   {\label{fig:summary} {\it Above:}
The number of detector pairs for each observing year. (In 
parenthesis is the number of pairs used in the current CMB analysis.)
In addition to bolometers behind feedhorns, along the perimeter of the 
focal plane are ``dark" bolometers not coupled to radiation, un-etched 
modules used as thermistors, and fixed resistors for diagnostic purpose.\\
{\it Right:} Nominal 2007/2008 layout of the \bicep\ beams, with six sections having 
alternating ``Q" or ``U" detector orientations with respect to the radial 
vector from the center.  The observations are performed with the focal 
plane orientations of $-45\deg, 0\deg, 135\deg,$ and $180\deg$, providing 
two independent and complete Q/U coverages of 
the field.}
\end{minipage} 
\begin{minipage}[t]{0.55\linewidth}
\vspace{-0.5in}
\begin{center}\begin{tabular}{c}
\includegraphics[trim=25 165 80 140,clip, width=\linewidth]{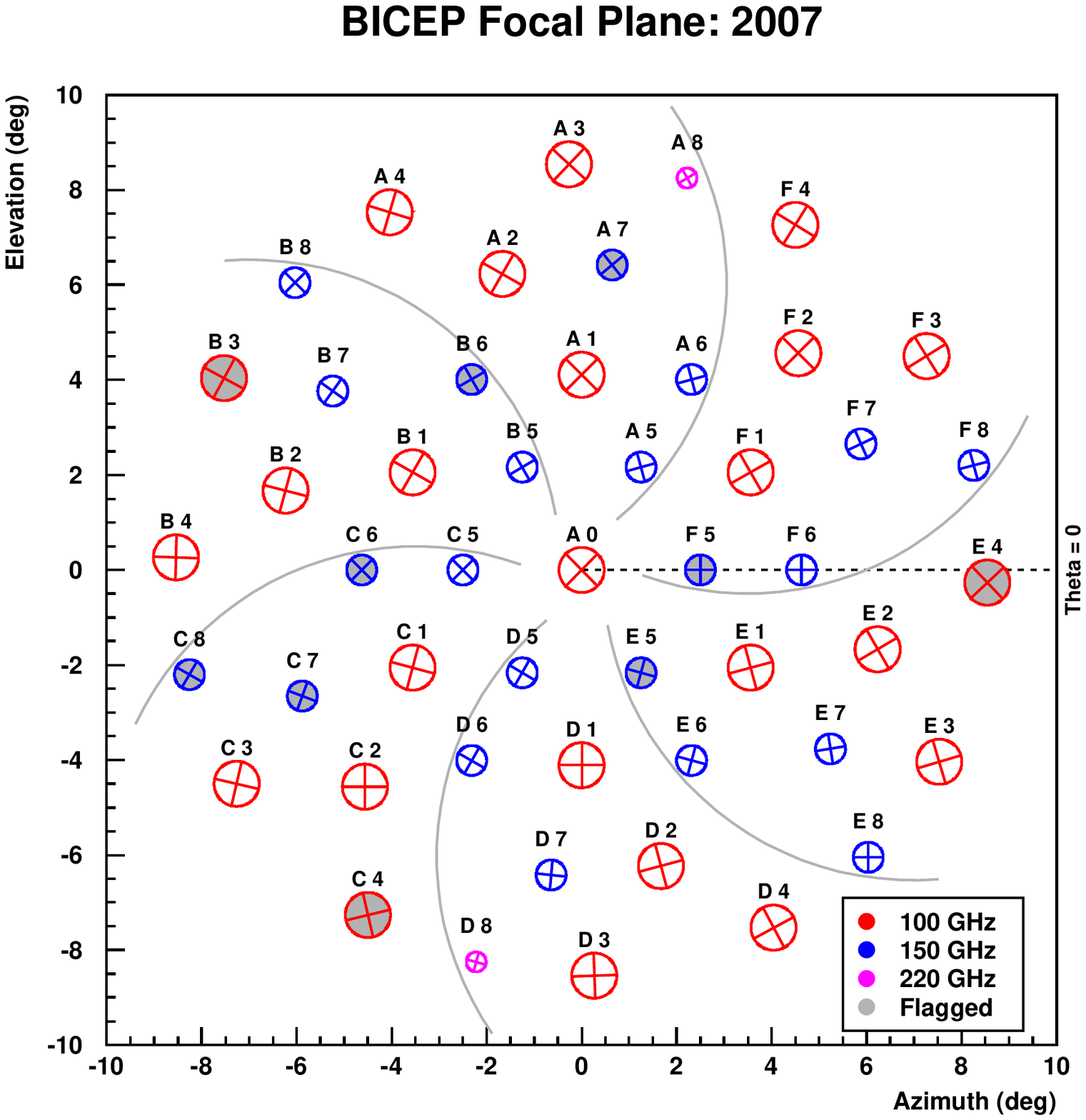}
\end{tabular}
\end{center}
\end{minipage}
\end{figure}

Figure~\ref{fig:summary} shows the number of detector pairs employed in 
each observing year, as well as the locations, orientations, and 
nominal full width half maximum (FWHM) of all the beams.
Between the first and second years, we added 220 GHz feedhorns in place of 
two of the 150 GHz ones along with the appropriate filters.  We also 
replaced four bolometers because of their slow temporal response, high 
noise level, or poor polarization efficiency.
We have kept \bicep\ cold and operating without interruption 
since this minor upgrade in December 2006.

\begin{figure}[phtb]
\begin{center} \begin{tabular}{c}
\includegraphics[width=\linewidth]{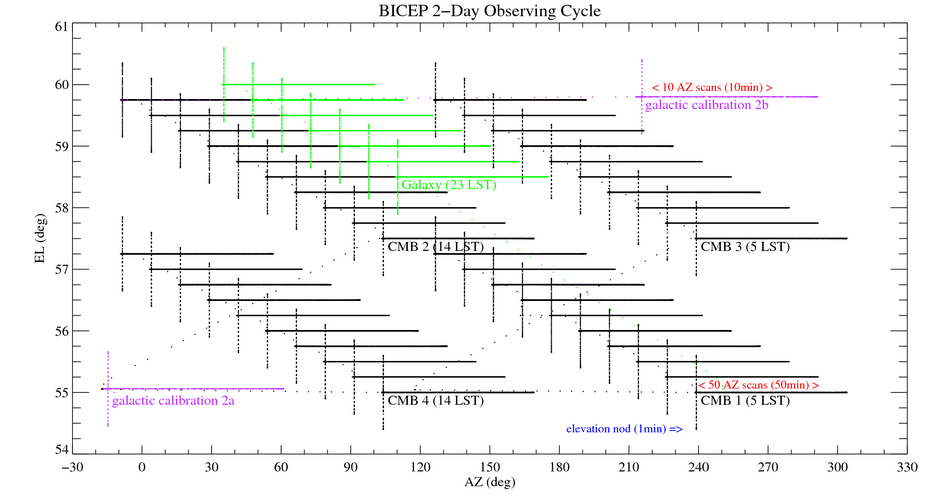}
\end{tabular} \end{center}
\caption[cycle]{\label{fig:cycle} A 48-hour observing cycle consists of 
6 hours for cycling the refrigerator (not shown), two 9-hour blocks of raster 
scans on our main CMB field on the first day (CMB 1 and CMB 2), a 6-hour block of similar raster scans on the 
Galaxy, and two more 9-hour blocks on the main CMB field on the second day (CMB 3 and CMB 4),
performed over the same azimuth range as on the first day but in the alternate coverage order
of the elevation range, allowing a test for ground contamination.
At the beginning and the end of 
each set of 50-minute azimuth scans, a $\pm0.6\deg$ elevation ``nod" is 
performed to measure relative gains of every bolometer.}
\end{figure}

With this instrument, we map an 800 deg$^2$ field daily by scanning 
a 60$\deg$ range in azimuth at 2.8$\deg$/s with hourly $0.25\deg$ 
steps in elevation.
Each two-day observing cycle (Figure~\ref{fig:cycle}) begins with 6 hours 
allocated for cycling the refrigerator, filling liquid nitrogen (every 2 
days) and liquid helium (every 4 days), measuring mount tilts, and 
performing star pointing calibrations.  The scan speed was selected to 
provide sufficient signal modulation against the $1/f$ atmospheric drifts 
while limiting motion-induced thermal fluctuations at the detectors in our 
science frequency band to sub-nK levels.

\section{INSTRUMENT CHARACTERIZATION}

To process the collected time streams into co-added polarization maps with 
systematic errors tolerable for our target sensitivity, a very accurate 
characterization of the detectors and their beams is crucial.
The voltage response of a PSB to radiation characterized by Stokes 
{$I,Q,U$} can be modeled as:
\begin{equation} \label{eq:model}
d(t) = H_t \otimes \frac{s}{2} \int {\rm d}\nu A_e F^\star_\nu \int {\rm d}\Omega\; 
P(\Omega) \left[ I + \frac{1-\epsilon}{1+\epsilon} (Q \cos{2\psi} + U 
\sin{2\psi}) \right]
\end{equation}
where $\psi$ is the orientation angle of the PSB's maximum response to linearly polarized
light, $\epsilon$ is the 
polarization leakage, $P(\Omega)$ is the beam function, $F^\star_\nu$ is 
the spectral response, $A_e$ is the effective antenna area, $s$ is the 
responsivity at 0 Hz, and $H_t \otimes$ signifies the convolution of the 
time-domain impulse response associated with the detector's frequency 
transfer function.

The requirement for the accuracy in the specification of these 
instrumental quantities were guided by simulated observations of the CMB 
sky to quantify the systematic errors that would induce false B-mode 
signal at the level of $r$=0.1. 
This benchmark level corresponds to B-mode fluctuations of 
$\sim$0.1$\muK$ rms at degree scales.
Table~\ref{tab:systematics} summarizes the instrument properties and
$r$=0.1 benchmark levels for their characterization,
as well as the measured results described in this section.
Each instrument property has been characterized to a level
of precision adequate for B-mode measurements at the $r$=0.1 level.



\begin{table}[!htb]
\small
\begin{center}
\caption{\small Potential Systematic Errors for \bicep.}
\label{tab:systematics}
\begin{tabular}[c]{|l|cc|}
\hline
Instrument properties           & Benchmark ($r$=0.1)   & Measured \\
\hline
Relative gain error: $\Delta(s_1-s_2)/s$           & 1.0\%            & ~~~0.4\% \\
Differential beam size$^a$: $(\sigma_1-\sigma_2)/\bar{\sigma}$  & 4.0\% &  $<0.2$\% \\
Differential pointing$^a$: $|\vec{r_1}-\vec{r_2}|/\bar{\sigma}$& 1.5\%  & ~~~~1.3\%$^b$ \\
Differential ellipticity: $(e_1-e_2)/2$         & 9.0\%         & $<0.1$\% \\
Polarization orientation error: $\Delta\psi$    & 4$\deg$       & 0.7$\deg$ \\
Polarized sidelobes to Galaxy$^c$               & - 8 dBi       & $<$ -38 dBi \\
Polarized sidelobes to ground$^c$               & -19 dBi       & $<$ -38 dBi \\
Cold-stage temperature stability$^d$: $\Delta$T & 1.3 nK        & $<0.5$ nK\\
Optics temperature stability$^d$: $\Delta$T$_{RJ}$ & 10 $\mu$K  & $<0.7$ $\mu$K\\
%
%
\hline
\multicolumn{3}{l}
{\scriptsize $^a$ $\bar{\sigma} =$ {\it FWHM}$/\sqrt{8\ln(2)}$.} \\
\multicolumn{3}{l}
{\scriptsize $^b$ A differential pointing which averages 1.3\% has been 
repeatably characterized to 0.4\% precision.} \\
\multicolumn{3}{l}
{\scriptsize $^c$ At 30$\deg$ from the beam center, based on the measured 
upper limit of 20\% (-7 dB) polarized response in the sidelobes.} \\
\multicolumn{3}{l}
{\scriptsize $^d$ Scan-synchronous, over $\ell = 30-300$.} \\
%
%
\end{tabular} \end{center} \end{table}

\subsection{Detector temporal transfer function}

Analysis of the time ordered data from each detector begins by 
deconvolving the temporal response using the measured frequency-domain 
optical transfer function of the detector.  Since the transfer function 
is proportional to the gain of the detector at each frequency, it 
directly affects the relative gains of a PSB pair to be differenced.  
The relative transfer functions must thus be measured with errors well below the 
1\% benchmark set for the relative gains.

At the nominal scan speed of $2.8\deg$/s in azimuth at $\sim$60$\deg$ 
elevations, our target angular scales of $\ell = 30-300$ fall 
into the frequency band of approximately 0.1--1 Hz.  Since the elevation 
nods described in Section~\ref{sec:relgains} probe the relative gains at 
$\sim$0.02 Hz, the transfer functions were measured down to 0.01 Hz.

\begin{figure}[!htb]
\begin{minipage}[t]{0.42\linewidth} \centering
   \includegraphics[width=\linewidth]{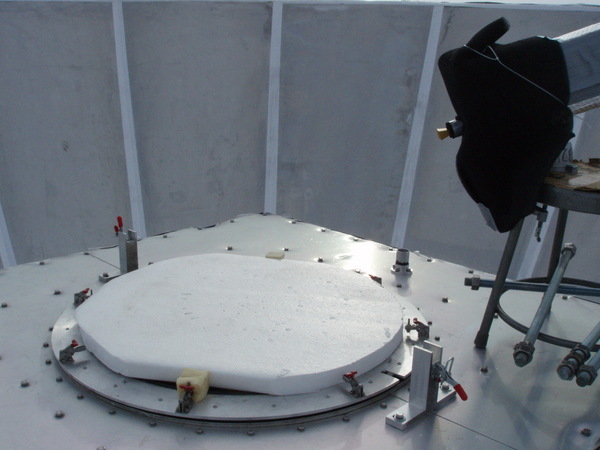}
   \caption[scatter] 
   {\label{fig:scatter} A setup in front of the telescope aperture for 
measurement of transfer functions.  Metal washers are embedded in the 
transparent zotefoam sheet to scatter the PIN switched signal into the 
beam while keeping the total loading similar to that during nominal CMB 
observations.}
\end{minipage} \hfill
\begin{minipage}[t]{0.56\linewidth} \centering
   \includegraphics[trim=0 0 0 20, clip, width=\linewidth]{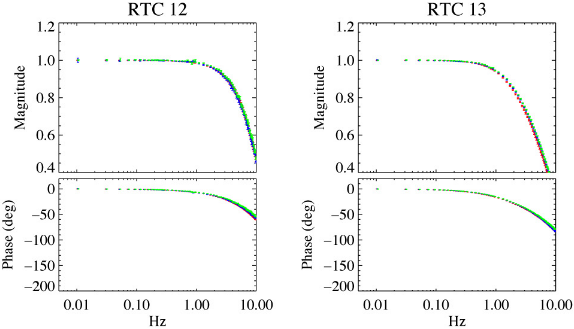}
   \caption[transfer]
   {\label{fig:transfer} Measured transfer functions with error bars for a 
typical pair of PSBs under different loading conditions and signal 
magnitudes: low-loading/high-signal (red), low-loading/low-signal
(purple), high-loading/high-signal (blue), and high-loading/low-signal
(green).}
\end{minipage}
\end{figure}

The primary measurement technique involved analyzing the step response to 
a fast-switching square-wave source (Gunn oscillator or broadband noise 
source) operating at 0.01--0.1 Hz, while under optical loading conditions 
representative of CMB observations (Figure~\ref{fig:scatter}).
Possible dependence on background loading and detector non-linearity were 
explored by repeating the measurement with extra loading from sheets of
emissive foam placed in the beam and/or with different signal strengths.

Figure~\ref{fig:transfer} shows the measured transfer functions for a 
representative PSB pair.
The relative gain mismatch due to measurement errors is found to be 
$<$0.3\% over the frequency range of 0.01--1 Hz.
Although the measured transfer functions fit the following model,
\begin{equation} \label{eq:transfer}
{\tilde H}(\omega) \propto \frac{1-\alpha}{(1-i\omega\tau_1)(1-i\omega\tau_2)}
+ \frac{\alpha}{(1-i\omega\tau_\alpha)},
\end{equation}
their signal-to-noise ratio was high enough that they could be directly 
inverse Fourier transformed to define the deconvolution kernels.
The median time constants were $\tau_1 \sim 20$ ms and $\tau_2 \sim 5$ ms.
From the first year, 6 channels at 150 GHz were excluded from CMB analysis 
due to excessive low frequency roll off. 
Two of the worst were in a PSB pair and 
were replaced at the end of the year.

Between the two years, the transfer function measurements generally agreed 
to within 0.5\% rms across the signal band.  Two exceptions were excluded 
from the first year CMB data, since the second year measurements were 
more reliable at the lowest frequencies.
Details of the measurements and analysis are in Yoon 2007 \cite{Yoon2007}.

\subsection{Beam widths and pointing}

Ideal differencing of a PSB pair is limited by any mismatch in the beams 
that can turn unpolarized signal into false polarization.  The differences 
between two nearly circular beams can be categorized into three dominant 
types: differential size, pointing, and ellipticity.  The beam size and 
ellipticity differences are sensitive to the second derivative of the 
temperature field, while differential pointing is also sensitive to the 
temperature gradient.  If the direction and magnitude of differential 
pointing can be established, the resulting leakage of CMB temperature 
gradients into polarization can in principle be estimated and accounted 
for.

\begin{figure}[!htb]
\begin{minipage}[t]{0.55\linewidth} \centering
 \includegraphics[trim=85 20 100 40, clip, width=\linewidth]{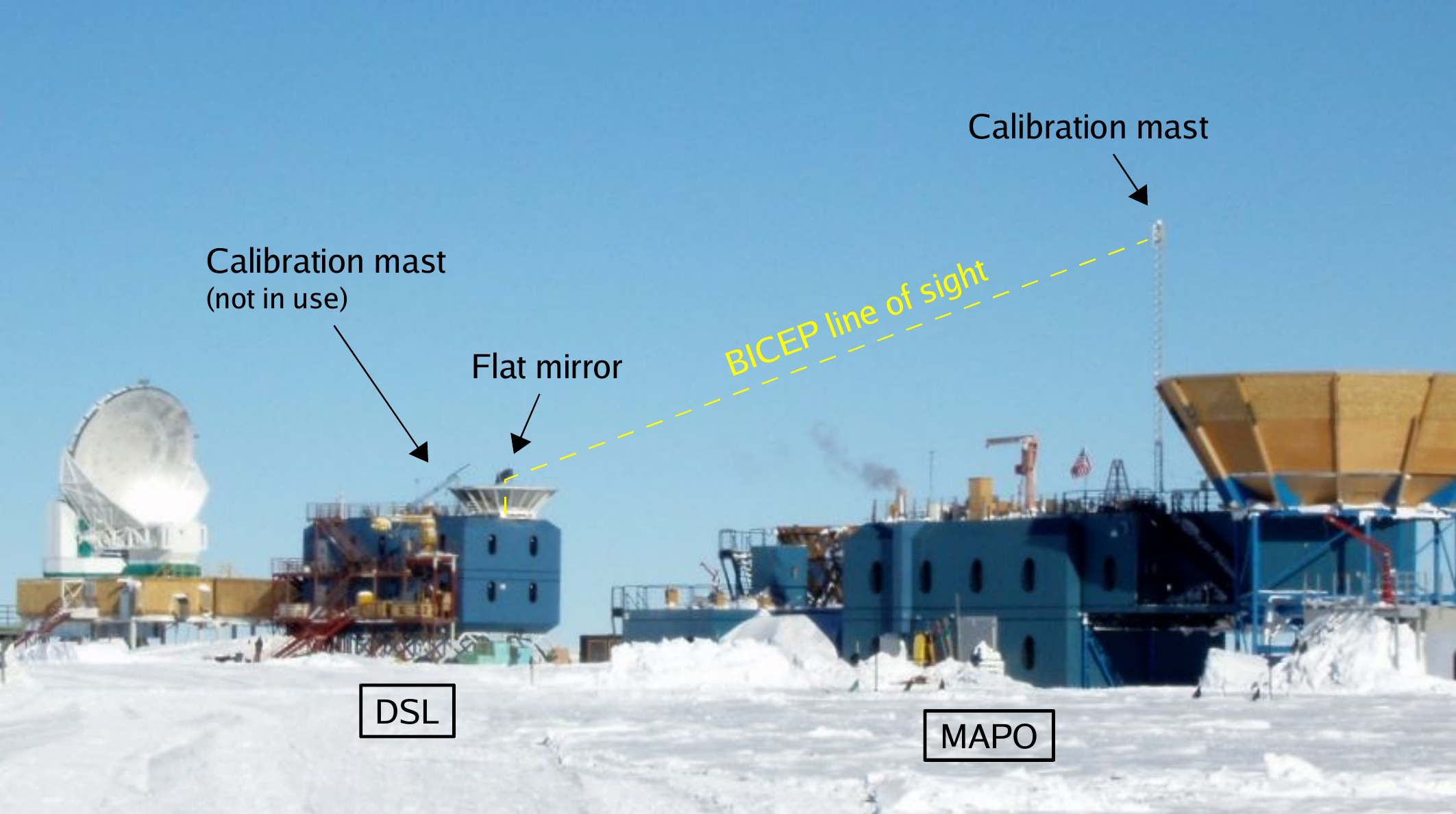}
 \caption[beam_setup]
 { \label{fig:beam_setup} The beam mapping setup on site consisted of 
sources mounted on top of fold-over masts.  When using the mast on the 
MAPO building (200 m from the Dark Sector Laboratory), a flat mirror is 
mounted to direct the beams over the ground screen.}
\end{minipage} \hfill
\begin{minipage}[t]{0.44\linewidth} \centering
 \includegraphics[width=\linewidth]{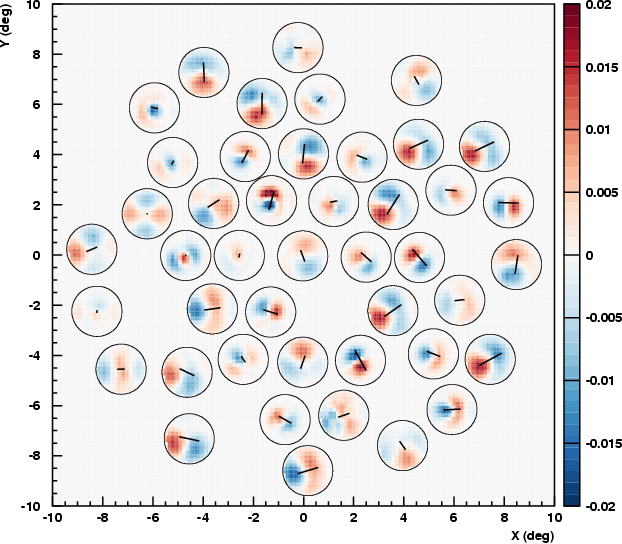}
 \caption[composite_diff] 
 { \label{fig:composite_diff} Beams for each PSB pair are
normalized and differenced to produce this composite differential beam
map.  The overplotted lines show the fitted centroid offsets magnified by 
a factor of 100.}
\end{minipage}
\end{figure}

The beams were mapped by raster scanning a bright source at various 
telescope orientations about the boresight.
The far field of the telescope is about 50~meters from the aperture, which 
permitted measurements in a high bay prior to telescope deployment and with the 
instrument installed at the South Pole.
In the high bay, a thermal blackbody source was used at a 40-meter 
distance, consisting of a liquid nitrogen load behind chopper blades 
covered with ambient temperature absorber.
At the South Pole, a temporary mast was installed on the rooftop 
outside of the ground screen, allowing us to position a source at 
60$\deg$ elevation 10~meters away.
For a truly far-field measurement, an additional mast was installed on 
the roof of the Martin A. Pomerantz Observatory (MAPO), at a distance of nearly 
200~meters, and a 
flat mirror was temporarily mounted above the telescope to direct 
the beams down to the low elevation of the mast 
as well as to low elevation astronomical sources
(Figure~\ref{fig:beam_setup}).  The sources used included an ambient 
temperature chopper against the cold sky, a broadband noise source, the 
Moon, and Jupiter.

The measured beams are well fit with a Gaussian model, typically resulting 
in 1\% residuals in 
amplitude. The fitted centroids are repeatable to about 0.02$\deg$, 
although the accuracy of the absolute locations is currently limited by 
uncertainties in parallax corrections.
The average measured FWHMs are 0.93$\deg$ and 0.60$\deg$ for 100 and 150 
GHz, respectively, about 5\% smaller than predicted from physical optics 
simulations.  The beams have small ellipticities of $<$1\% for 100~GHz and 
$<$1.5\% for 150~GHz.

A composite of the differenced normalized beams for each PSB pair is shown 
in Figure~\ref{fig:composite_diff}, measured with the broadband noise 
source, an amplified thermal source that is ideal for probing low-level 
effects.  The largest beam mismatch effect is a pointing offset that gives 
rise to dipole patterns in many of the differenced beams.  The median 
differential pointing offset is 0.004$\deg$, or $\sim$1\% of the beam size 
$\bar{\sigma}$, and the offsets were repeatable between observations 
performed at different telescope orientations of both the noise source and
the Moon to within the measurement uncertainty, 0.4\% of $\bar{\sigma}$.  
The measurements of 
differential size and ellipticity are less repeatable, but the upper 
limits of 0.2\% and 0.1\%, respectively, are negligibly small.

Finally, to co-add maps made with different PSB pairs, the actual 
locations of all the beams must be determined.  
This was accomplished by first making a full season co-added map of the 
CMB using the design locations and then cross-correlating the temperature 
anisotropy pattern with single detector maps to adjust the individual 
beam coordinates.  These adjusted coordinates were then used to iterate 
this process, resulting in derivation of the absolute beam locations 
with an uncertainty of 0.03$\deg$ rms, based on the agreement between the 
first and second years.  For \bicep, using the CMB temperature 
fluctuations proved more effective than attempting a similar procedure 
with Eta Carinae, the brightest compact source accessible.

\subsection{Polarization orientations and efficiencies}


Angles of the PSBs can vary from their design orientations due to the mechanical 
tolerances with which they were mounted.  The deviation from 
perfect orthogonality of a pair simply reduces its efficiency for 
polarization; however, an error in the overall orientation of the pair can 
lead to mixing of E-modes into B-modes.  To ensure this mixing is well 
below the level of the B-mode spectrum for $r$=0.1, the calibration 
procedure was designed to determine the polarization orientations to 
within $\pm$1$\deg$.

Another factor, though less important, is that the PSBs are not perfectly 
insensitive to polarization components orthogonal to their orientations, 
effectively reducing the polarization efficiency to 1$-\epsilon$.  To 
achieve 10\% accuracy in the amplitudes of the polarization power 
spectra, our goal was to measure polarization leakages $\epsilon$ to 
within $\pm$0.02.

\begin{figure}[!htb]
\begin{minipage}[t]{0.32\linewidth}
   \centering\includegraphics[width=\linewidth]{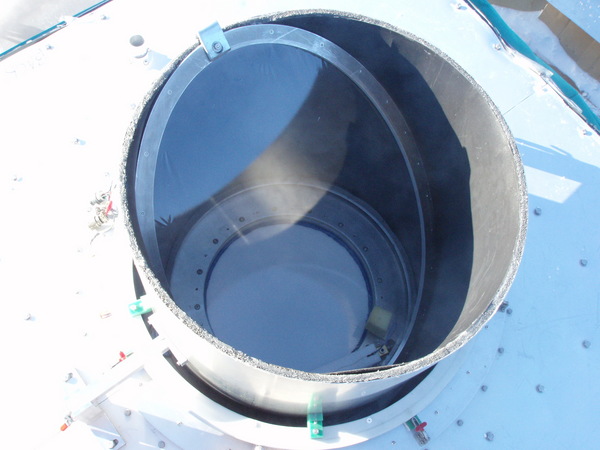}
   \caption[calibrator]
   {\label{fig:calibrator} The dielectric sheet calibrator for measuring $\psi$
consists of a beam-filling polypropylene sheet and a highly emissive black 
lining as an ambient load, injecting partially polarized radiation into the 
telescope aperture.  The device is mounted on the azimuth stage, which can 
rotate about the telescope's boresight when pointed at zenith.}
\end{minipage}\hfill
\begin{minipage}[t]{0.32\linewidth}
   \centering\includegraphics[width=\linewidth]{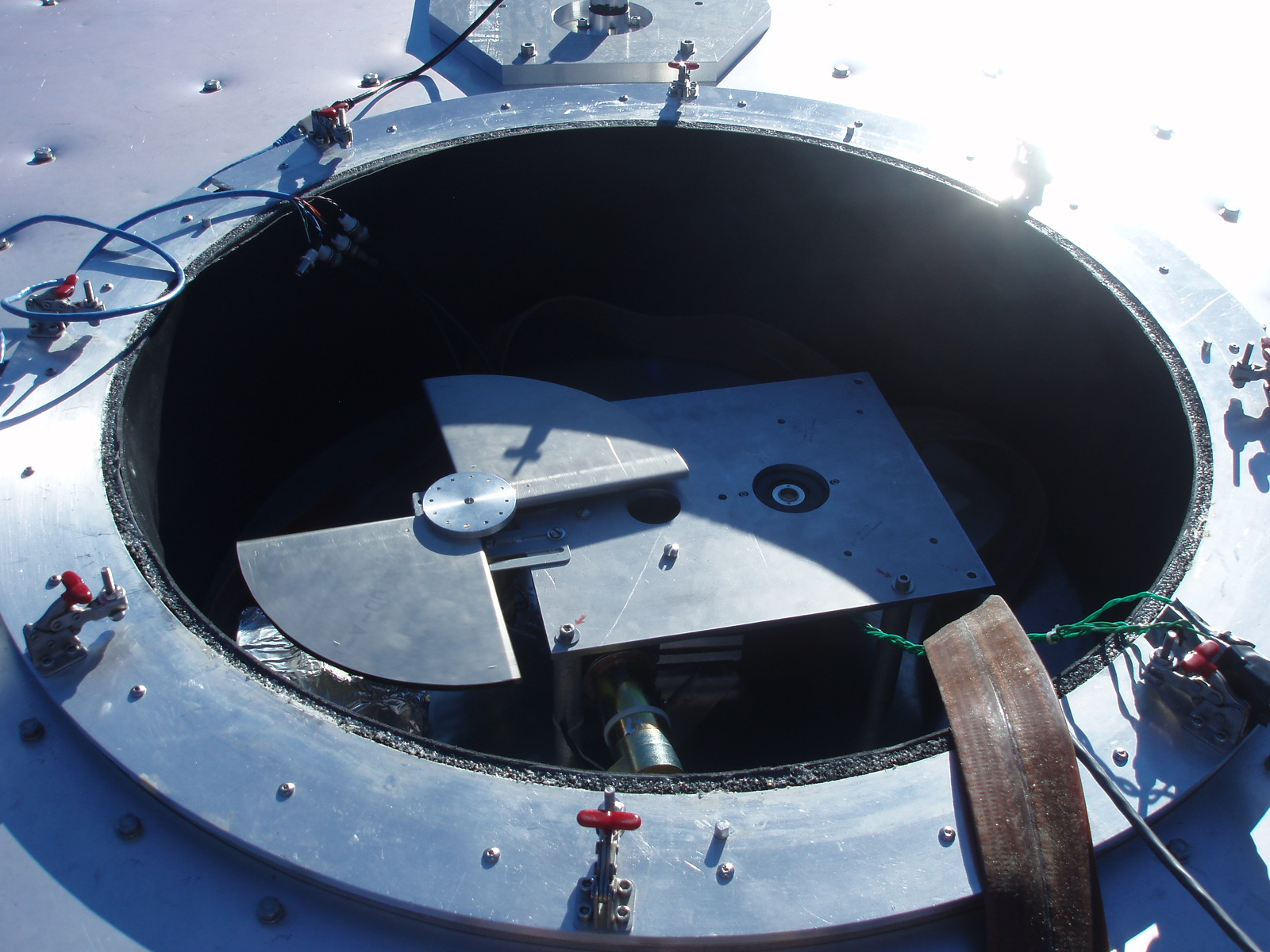}
   \caption[aperture]
   {\label{fig:aperture} A device above the cryostat window for measuring 
$\epsilon$ (and $\psi$).  The window is covered with a metal plate with a 
2-cm Eccosorb aperture, and a 10-cm diameter wire grid is on a rotation 
stage under the circular aperture of the rectangular plate.  The chopper 
modulates the load between the ambient temperature and the cold sky.}
\end{minipage}\hfill
\begin{minipage}[t]{0.32\linewidth}
   \centering\includegraphics[width=\linewidth]{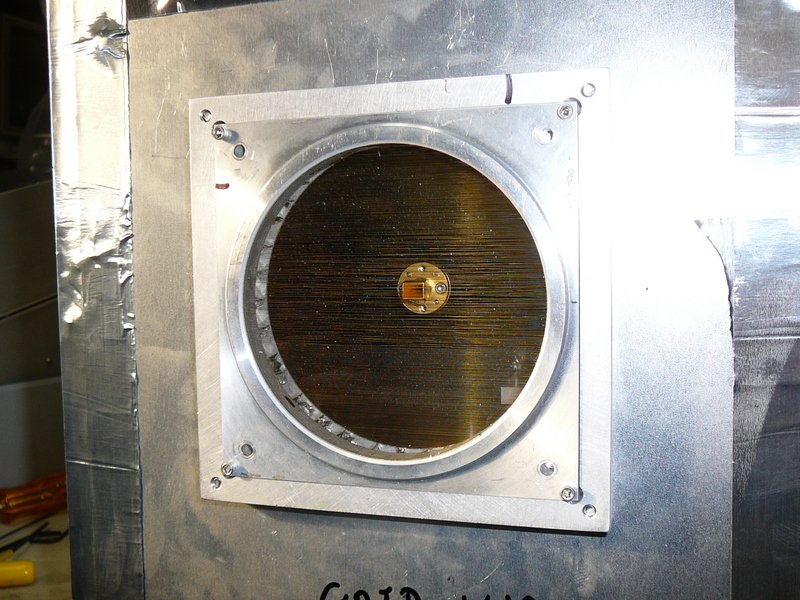}
   \caption[bns]
   {\label{fig:bns} Another calibration source used on top of a mast for 
measuring $\epsilon$ (and $\psi$).  The broadband noise source at 100 or 150 
GHz outputs power through the rectangular feedhorn oriented for either 
vertical or horizontal polarization and through a precisely aligned wire grid to minimize 
cross-polarized signal.}
\end{minipage}
\end{figure}

As described in the 2006 Paper, the polarization orientations were 
measured using a rotatable dielectric sheet device shown in 
Figure~\ref{fig:calibrator}.  The measurements were performed several 
times throughout each observing year, and produced repeatable results for 
the individual PSB orientations with 0.3$\deg$ rms.  The PSB pairs were 
orthogonal to within 0.1$\deg$, and together were within $\pm$1$\deg$ of 
the design orientations shown in Figure~\ref{fig:summary}.  The absolute 
orientation was found to within $\pm$0.7$\deg$, limited by the 
accuracy in determining the dielectric sheet orientation relative to the 
azimuthal orientation of the entire array.

This absolute orientation was confirmed by two other independent methods 
that were primarily used to measure the polarization leakages.  One method 
employed a rotatable wire grid on top of the telescope aperture, with a 
chopper modulating the completely polarized load between the ambient 
absorber and the cold sky (Figure~\ref{fig:aperture}).  Fitting a 
sinusoidal curve to the PSB response as a function of the wire grid angle 
gives the polarization efficiency and orientation.  The measurements for 
both years gave polarization leakages within $\pm$0.02 of each other, with 
a median value of $\epsilon$=0.05.

The other method used a modulated broadband noise source with a 
rectangular horn behind a wire grid, mounted on the mast 200 meters away 
(Figure~\ref{fig:bns}).  This source was raster scanned by each of our 
beams with 18 different detector orientations with respect to the wire 
grid, fitting a 2-dimensional Gaussian to each raster.  The measured 
leakages were slightly lower with a median of $\epsilon$=0.04, and the 
uncertainty was $\pm$0.01 based on the scatter against the results from 
the first method.  
One 150 GHz bolometer was flagged for having an $\epsilon$$>$0.12 and was 
replaced at the end of the first year.

\subsection{Spectral response}

As described in Section~\ref{sec:relgains}, relative gain calibration 
used to remove unpolarized CMB and atmospheric emission is based on 
elevation nods that subject the bolometers to varied atmospheric loads.  
Because the CMB and the atmospheric emission have different spectral 
shapes, the spectral response of the PSB pair must match well enough to 
prevent errors in the relative responsivity to the CMB.

\begin{figure}[!htb]
\begin{minipage}[t]{0.48\linewidth}
 \centering
 \includegraphics[width=\linewidth]{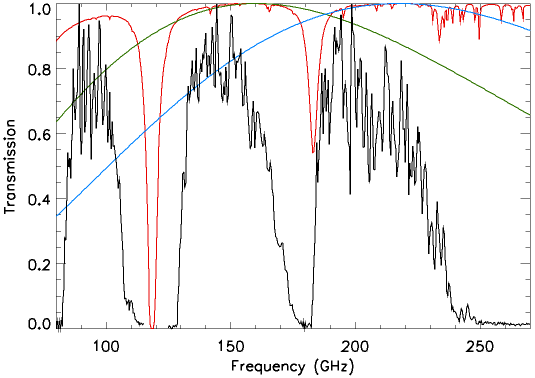}
 \caption[spectra]
 { \label{fig:spectra} The average measured spectral response for each 
of \bicep's frequency bands, normalized with respect to the maximum.
Overplotted are the atmospheric transmission at the South Pole (red), the 
CMB spectrum (green) and its temperature derivative (blue).}
\end{minipage} \hfill
\begin{minipage}[t]{0.48\linewidth}
 \centering\includegraphics[width=\linewidth]{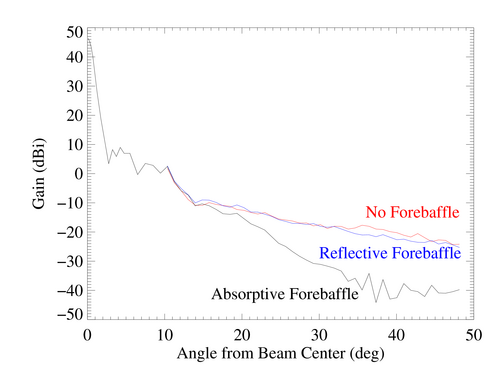}
 \caption[sidelobe] 
 { \label{fig:sidelobe} The far sidelobe response for 
the central feed.
Being cylindrical, a forebaffle with a bare aluminum surface had little 
effect on the sidelobe rejection, while the addition of the absorptive 
lining (Eccosorb HR-10) provided up to an additional 15 dB attenuation.  
When the telescope is at its lowest elevation of 50$\deg$, the lip of the 
outer ground screen is $\sim 30\deg$ from the beam center of the central 
feed.}
\end{minipage}
\end{figure}

The spectral response of each channel was measured using two separate polarized 
Fourier 
Transform Spectrometers with a maximum resolution of 0.3 GHz.  Within each 
frequency band, the spectra were very similar from channel to channel 
(average spectra shown in Figure~\ref{fig:spectra}), and the preliminary 
upper limit on the expected differential gains due to the spectral 
mismatch appears to be at an acceptable level.

In addition to the main band, we verified that there is no significant 
response at higher frequencies due to leaks in the low-pass filter 
stacks.  Thick grill filters with cut-off frequencies of 165 GHz and 255 
GHz were used at the telescope aperture one at a time and the response to 
a chopped thermal source was measured.  150 GHz channels showed no sign of 
leaks down to the noise floor at -35 dB, while 100 GHz channels exhibited 
leaks at $\sim$ -25 dB level somewhere beyond 255 GHz.  The magnitude of 
this small ($\sim$0.3\%) leak was consistent between the PSBs in each 
pair, so that possible effect on relative responsivities is expected to be 
negligible.

\subsection{Far sidelobe rejection}

The design criterion for the ground shields was to reject the ground 
radiation to a level below our target B-mode polarization sensitivity.  
Polarized sidelobes can produce spurious signals by also coupling to 
emissions from the Galaxy and the outer ground screen.

The far sidelobe response of \bicep\ including the forebaffle (see 
Figure~\ref{fig:instrument}) was measured 
using a modulated microwave source (Gunn oscillator or broadband noise 
source) on top of the mast within a line of sight.  The telescope was 
stepped away in elevation up to 60$\deg$ away from the source in 0.5$\deg$ 
increments, making one revolution about the boresight at each step for a 
complete azimuthal coverage.  This measurement was performed with several 
source attenuations down to -50 dB to probe the many decades of gain with 
sufficient signal-to-noise ratio while also measuring the main beam 
without saturating the detector.

The sidelobe maps are azimuthally averaged to obtain a radial profile 
(Figure~\ref{fig:sidelobe}).
To evaluate the far sidelobe rejection performance, this level of 
response with a measured upper limit of 20\% polarization was convolved 
with a model map of the predicted dust emission from the Galaxy.  The 
resulting B-mode contamination in our CMB field was found to be 30 dB below the 
$r$=0.1 level.  
The same exercise was repeated for a conservative model of varying snow 
accumulation on the ground screen panels.
The contamination was larger, but the achieved rejection level was still $\sim$20 
dB better than the benchmark
(see Table 1).
Potential ground contamination can also be directly probed in our data through jackknife tests.
If necessary, our scan strategy allows us to separate it from the sky 
signal, although so far no ground signal subtraction has proven 
necessary.

\section{CALIBRATION}

In addition to relatively static instrumental properties that are 
characterized at most annually, possibly dynamic quantities are
calibrated routinely throughout the observing season.  These include 
the detector responsivities, especially the relative gains of the PSB 
pairs, and the telescope pointing parameters.

\subsection{Relative detector gains} \label{sec:relgains}

Polarization measurement with \bicep\ relies on PSB pair differencing, in 
which the relative gains within each pair must be accurately determined 
to prevent the bright unpolarized component from inducing false 
polarized signals.
Simulations suggest that the pair relative gains should be accurate to at 
least 1\% to limit the leakage of CMB temperature anisotropy into 
significant B-mode contamination.

\begin{figure}[!htb]
\begin{minipage}[t]{0.48\linewidth}
   \centering
   \includegraphics[trim=0 0 0 25, clip, width=\linewidth]{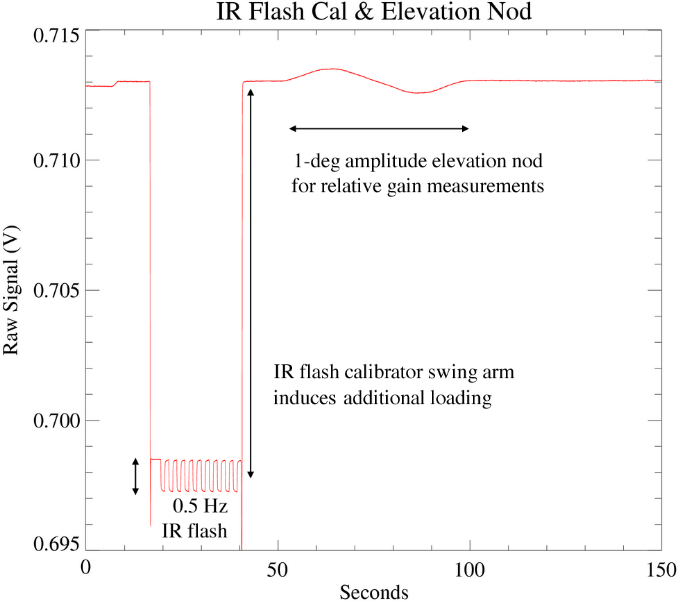}
   \caption[flashelnod]
   { \label{fig:flashelnod} A bolometer time stream during a relative 
responsivity calibration procedure.  An infrared source is swung into the 
beam to inject a signal of very stable amplitude, useful for tracking any 
gain variations.  The elevation nod uses the modulation of the atmospheric 
emission.\\} 
\end{minipage} \hfill
\begin{minipage}[t]{0.49\linewidth}
   \centering
   \includegraphics[trim=0 0 0 27, clip, width=\linewidth]{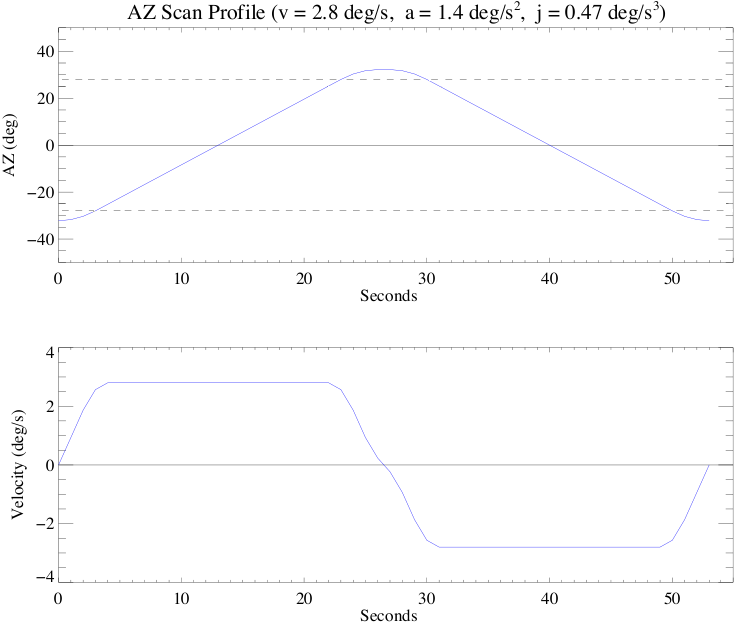}
   \caption[azscan]
   { \label{fig:azscan} A 2.8$\deg$/s azimuth scan profile used in mapping 
the CMB field.  The acceleration during turn-arounds is limited to 
1.4$\deg$/s$^2$; the jerk is limited to 0.47$\deg$/s$^3$.  A 20 second 
portion of mostly constant-velocity scan is used from each direction, 
resulting in a 75\% scan efficiency.\\}
\end{minipage}
\end{figure}

Relative gains are derived from elevation nods performed once at the 
beginning and once at the end of every one-hour constant-elevation scan 
set (Figure~\ref{fig:cycle}).  The elevation motion of the telescope 
during a nod is a rounded triangle wave with a peak-to-peak amplitude of 
$1.2\deg$ that injects a $\sim\pm$0.1~K optical loading signal 
(Figure~\ref{fig:flashelnod}).  The bolometer responses are fit 
to a simple air mass model of atmospheric loading versus elevation 
($T_{atm} \propto$ 1/sin($EL$)) to derive the relative gains across the 
array.

The elevation motion generates thermal disturbances on the focal plane, 
although they are limited by performing the nod slowly over a 50 second 
period.  To reduce the effect of the thermally-induced false signals, the 
two elevation nods for each scan set are performed in opposite patterns 
(up-down-return and down-up-return) and the average gain is used.  While 
the two patterns result in a small systematic difference in the 
individual gains, the pair {\it relative} gains are consistent to within 
$\pm$0.3\% with no systematic difference between the two patterns.  Over 
a time scale of months, the measured relative gains are stable with 
$\sim$1\% rms and exhibit no systematic variation with the optical 
loading level.  As a cross check, relative gains have also been 
derived from correlating time-stream atmospheric fluctuations within PSB 
pairs, and the results agree with the elevation nods within $\pm$0.3\%.

The common-mode rejection was ultimately assessed by cross-correlating
individual PSB pair-sum and pair-difference maps to quantify the level of
leakage of the CMB temperature anisotropy into pair-differences.  The median
correlation over the PSB pairs was 0.4\% over the angular scales of 
interest, validating the differencing technique to this level.

The infrared flash calibrator described in the 2006 Paper gave a very
repeatable (0.2\% rms) response between the beginning and the end of the
one-hour scan sets and also showed that the individual gains are stable with
1\% rms across the full elevation range.  However, there are several 
uncertainties associated with the calibrator, including $\sim$3~K excess 
optical loading introduced by the swing arm, and unknown polarization of 
the infrared source.  The relative gains from the flash calibrator have 
therefore not been used in the analysis.

Assuming that the average responsivity to the CMB of all the detectors is 
constant, the measured individual gains are scaled for each 
1-hour scan set such that the mean gain over each frequency band is 
constant.  The gain-adjusted time streams are co-added to form CMB maps in 
detector voltage units, and the \bicep\ and WMAP temperature maps 
(identically smoothed and filtered) are cross-correlated to derive 
absolute gains for each of \bicep's frequency bands.  The absolute gains 
are consistent with those obtained from the dielectric sheet calibrator, 
which have 10\% uncertainty.

\subsection{Telescope pointing}

Pointing errors greater than 1\% of the beam size $\bar{\sigma}$ could
contaminate polarization at the $r$$\sim$10$^{-4}$ level
\cite{2003PhRvD..67d3004H}.  Although this effect is far below our current
sensitivity target, our star pointing camera (described in the 2006 Paper) was
developed to achieve this 1\% accuracy, which for \bicep\ corresponds to
10$^{\prime\prime}$.

The telescope pointing is established using star observations with an optical
camera rigidly bolted to the top surface of the cryostat.  Pointing
calibrations are performed every two days during the refrigerator cycles,
weather permitting, as well as before and after each mount re-leveling.
In each calibration run, the telescope points at 24 stars
(down to magnitude 3) at boresight orientation angles of -45$\deg$, 45$\deg$,
and 135$\deg$, and the azimuth and elevation offsets required to center the
stars are recorded.  The pointing data are fit to a 10-parameter model with
typical residuals of 10--12$^{\prime\prime}$ rms.
The pointing model has been checked by cross-correlating the CMB 
temperature anisotropy patterns between the pointing-corrected daily maps 
and the cumulative map; no systematic offsets or drifts were detected.

The tilt of the telescope mount is monitored every two days with two
orthogonal tilt meters mounted on the azimuth stage.  We have observed seasonal
tilt changes of up to 0.5$^\prime$ per month, possibly due to the
building settling on the snow, and generally have re-leveled the mount before the tilt
exceeds 1$^\prime$.

\section{PERFORMANCE}

We began our telescope installation work at the South Pole in November 2005,
cooled down the cryostat in December, and captured first astronomical light
a month later.  Following calibration measurements and tests of the 
observing strategy, \bicep\ began CMB observations in February 2006.  
The instrument has since operated nearly continuously and will continue 
observing in its current configuration until the end of 2008.


\subsection{Observing efficiency}

Excluding any incomplete 9-hour blocks, \bicep\ acquired 180 days of CMB
observations during 2006, 
in which a significant fraction of the observing season
was devoted to calibration measurements.  The amount of CMB observations
increased to 245 days in 2007.  CMB data acquired during the Austral
summer are generally of lower quality than the winter data because of several
factors, including mediocre weather conditions and increased station 
activities.  Although \bicep\ is capable of observing during the 
summer---there has not been evidence for Sun contamination---we restrict our CMB 
analysis to data taken during February--November.

The first 2.5 months of data in 2006 were excluded from the current 
analysis because a different scan strategy was being investigated at this 
time, and a small but measurable level of radio frequency (RF) 
interference was detected in the warm receiver electronics.  The RF 
interference was successfully eliminated by carefully sealing the RF 
shield surrounding the electronics.  As a coarse weather cut, we have 
excluded 9-hour blocks if the relative gains derived from elevation nods 
have a standard deviation above 20\%, averaged over the channels.  After 
these cuts, 116 days in 2006 and 226 days in 2007 remain for our baseline 
CMB analysis.

The data set is slightly reduced due to the occasional presence of snow on the
window.  Infrared images of the telescope window were captured every hour, and
7\% of the scan sets in 2006 and 2\% in 2007 were flagged for visible snow
accumulation.  Furthermore, 3\% of PSB pair time streams are flagged due to
cosmic ray hits, glitches, and $>$3\% mismatch in relative gain measurements 
from elevation nods over one-hour periods.  Accounting for the 75\% scan 
efficiency (Figure~\ref{fig:azscan}) and the scheduled calibration routines, the 
net CMB observing efficiency is 60\% during the CMB observing blocks and 45\% 
overall during each two-day cycle.


\subsection{Thermal stability}

Temperature drifts in the focal plane and optics can produce
artifacts if they have a scan-synchronous component.
The refrigerator cools to 233--236 mK, and the temperature of the focal 
plane is stabilized at approximately 250~mK using a 100 k$\Omega$ resistor 
as a control heater (nominally depositing $\sim$0.1 $\mu$W) in a PID 
feedback loop with a sensitive NTD Germanium thermistor.
During the first year, the thermal control scheme used a thermistor on the 
focal plane closest to the thermal strap connected to the refrigerator.  
Since motion dependent thermal disturbances were observed, and they 
presumably transmit through the strap, the control scheme for the 
subsequent years used two new pairs of thermistors mounted on the strap 
itself immediately adjacent to the control heater.  
In addition, to better suppress any microphonically induced heatings, the 
thermal straps were further reinforced using Vespel supports.
Along with an increased response speed of the control loop, the spectra of 
the focal plane thermistors improved visibly from the first year to the 
second.

The temperature stability of the focal plane was found to vary with 
azimuth scan speed and the telescope orientation about its boresight.
The stability was investigated under a variety of telescope operating 
conditions, including scan speeds in a range of $1.0\deg$/s -- 
$4.0\deg$/s, and 16 evenly spaced boresight orientation angles.
We selected our 2.8$\deg$/s operating scan speed and four boresight angles
$\{-45\deg, 0\deg, 135\deg, 180\deg\}$ by minimizing the variance of the
scan-synchronous thermistor signals.

The bolometers' responsivities to the bath temperature were measured by 
fitting the output voltage to a 10~mK drop at the end of each refrigerator 
cycle.  The median thermal responsivity is 0.8~$\mu$K$_{\rm CMB}$/nK$_{\rm 
FP}$, and the median mismatch within PSB pairs is 0.08 $\mu$K$_{\rm 
CMB}$/nK$_{\rm FP}$.  In order to meet the $r$=0.1 target at 
$\ell\sim$100, thermal instabilities in the focal plane must be controlled 
to better than 1.3~nK rms.  The measured level of thermal fluctuations in 
\bicep\ is 0.5~nK rms and flat over the frequency range corresponding to 
$\ell=30-300$.

Since emission from \bicep's optics is expected to be 
unpolarized, the main concern with optics temperature drifts is in 
mis-calibration of PSB pair optical relative gains, which are controlled 
to better than 1\%.  Scan-synchronous fluctuations averaged over all the
individual bolometers and over a 2 month period showed only 
0.7~$\mu$K$_{RJ}$ variation, integrated over the relevant frequency range.




\subsection{Preliminary Results}

\begin{figure}[!htb] 
\begin{center}
\includegraphics[width=6.0in]{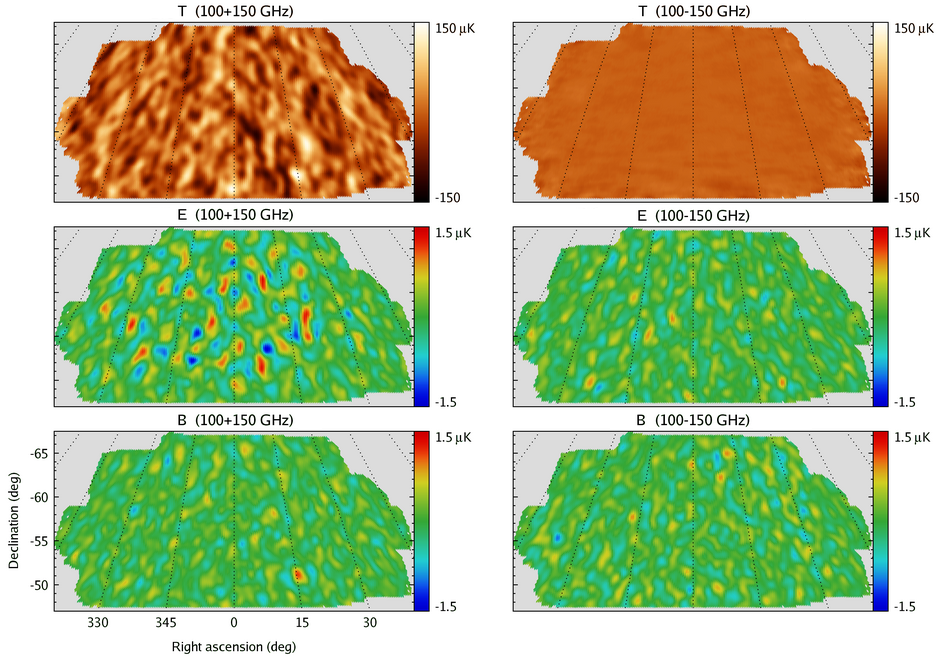}
\end{center} 
\caption{\label{fig:bicep_maps} 
Maps of the first two years of data, including frequency differnce maps 
(right column).
A 3rd-order polynomial is removed from each half-scan, and the maps 
are smoothed to 1$\deg$ resolution.
\emph{Top row:} CMB temperature anisotropy.
The faint striping in the difference map is due to residual atmospheric 
noise, which is removed by PSB differencing.
\emph{Bottom rows:}
Stokes Q and U maps are combined to produce Wiener-filtered E- and B-mode 
polarization maps.  The E signal map shows degree-scale structure at the 
expected level, while the other three maps are consistent with noise. 
}
\end{figure}

Temperature and polarization maps from \bicep's first two years of  
operation, as well as frequency difference maps formed by subtracting maps 
made at 100 and 150 GHz, are shown in Figure~\ref{fig:bicep_maps}.  
The CMB temperature anisotropy has been measured with high 
signal-to-noise, as demonstrated by the absence of structure in the 
difference map.  The faint striping, caused by residual atmospheric 
noise, is successfully removed by PSB differencing.


The E- and B-mode maps are calculated from apodized Stokes Q/U maps and 
are Wiener-filtered according to the expected E-mode signal divided by the 
beam function.  Degree-scale E-mode structure is visible, while the E and 
B difference maps and B signal maps are consistent with noise.  The 
100--150 GHz maps show no evidence of foreground contamination.

After 3700 hours of total integration time, the noise per 1 $\rm{deg}^2$ pixel
in the Q and U jackknife maps is measured to be 0.78~$\mu$K rms for 100~GHz
and 0.62~$\mu$K rms for 150~GHz, which is consistent with expectations.  The
map noise levels translate to a preliminary estimate of noise equivalent 
temperature (NET) per detector of 560~$\mu$K$\sqrt{s}$ at 100~GHz and 
430~$\mu$K$\sqrt{s}$ at 150~GHz.


\section*{ACKNOWLEDGMENTS}       

\bicep\ has been made possible through support from NSF Grant No. OPP-0230438, 
Caltech President's Discovery Fund, Caltech President's Fund PF-471, JPL 
Research and Technology Fund, and the late J. Robinson.  BGK gratefully 
acknowledges support from NSF PECASE Award \#AST-0548262.  We thank the 
South Pole Station staff for their continuing support and Steffen Richter 
for being a winter over we can rely on.
We thank our colleagues in \acbar, \boomn, \QUAD, Bolocam, and {\sc Spt}
for advice and helpful discussions, and Kathy Deniston for logistical 
and administrative support.


\bibliography{report}   
\bibliographystyle{spiebib}   

\end{document}